# Electron spin coherence exceeding seconds in high purity silicon

Alexei M. Tyryshkin[1], Shinichi Tojo[2], John J. L. Morton[3], Helge Riemann[4], Nikolai V. Abrosimov[4], Peter Becker[5], Hans-Joachim Pohl[6], Thomas Schenkel[7], Michael L. W. Thewalt[8], Kohei M. Itoh[2], and S. A. Lyon[1]

[1]Department of Electrical Engineering, Princeton University, Princeton, NJ 08544, USA

[2]School of Fundamental Science and Technology, Keio University, Yokohama, Kanagawa 2238522, Japan

[3]Department of Materials, University of Oxford, Oxford OX1 3PH, United Kingdom

[4]Institut für Kristallzüchtung, D-12489 Berlin, Germany

[5]Physikalisch-Technische Bundesanstalt, D-38116 Braunschweig, Germany

[6]VITCON Projectconsult GMBH, D-07745 Jena, Germany

[7]Lawrence Berkeley National Laboratory, Berkeley, CA, 94720, USA

[8]Department of Physics, Simon Fraser University, Burnaby, BC V5A 1S6, Canada

## Abstract

Silicon is undoubtedly one of the most promising semiconductor materials for spin-based information processing devices.[1,2] Its highly advanced fabrication technology facilitates the transition from individual devices to large-scale processors, and the availability of an isotopically-purified $^{28}$Si form with no magnetic nuclei overcomes what is a main source of spin decoherence in many other materials.[3,4] Nevertheless, the coherence lifetimes of electron spins in the solid state have typically remained several orders of magnitude lower than what can be achieved in isolated high-vacuum systems such as trapped ions.[5] Here we examine electron spin coherence of donors in very pure $^{28}$Si material, with a residual $^{29}$Si concentration of less than 50 ppm and donor densities of $10^{14-15}$ per cm$^3$. We elucidate three separate mechanisms for spin decoherence, active at different temperatures, and extract a coherence lifetime $T_2$ up to 2 seconds. In this regime, we find the electron spin is sensitive to interactions with other donor electron spins separated by ~200 nm. We apply a magnetic field gradient in order to suppress such interactions and obtain an extrapolated electron spin $T_2$ of 10 seconds at 1.8 K. These coherence lifetimes are without peer in the solid state by several orders of magnitude and comparable with high-vacuum qubits, making electron spins of donors in silicon ideal components of a quantum computer,[2,6] or quantum memories for systems such as superconducting qubits.[7-9]

Silicon has been recognized as a promising host material for spin-based electronic devices where information is stored and manipulated using the spin of the electrons, rather than their charge as in conventional electronics.[1,10] It is important to distinguish between two forms of information storage within spin: firstly the storage of *classical* information is possible using the orientation of the spin with respect to some externally applied or internal magnetic field (i.e. the 'spin-up' or 'spin-down' states). This forms the basis of spintronics, and corruption of the classical information can be characterized by the longitudinal electron spin relaxation time $T_1$. The electron spin is also capable of representing *quantum* information, using superposition states of spin-up and spin-down states with well-defined phase. This information is much richer than classical information, but often much more fragile as it requires the preservation of the full coherent spin state. The corruption of this phase information is characterized by the coherence lifetime $T_2$, which, though bounded by the relaxation time $T_1$, is often much lower due to additional mechanisms which only affect the spin coherence.

Very long $T_1$ values have been reported for electrons bound to shallow donors in silicon. For example, $T_1$ for phosphorous donors approaches an hour at 1.2 K and 0.35 T and shows a strong temperature dependence at higher temperatures (Figure 1). The temperature dependence of $T_1$ is well understood in terms of spin-phonon relaxation processes, including a one-phonon (direct) process, and two-phonon (Raman and Orbach) processes.[11,12]

The electron spin coherence times $T_2$ of shallow donors have also been studied previously, yielding times in the range of hundreds of microseconds to tens of milliseconds.[13-16] These times have prompted interest in donor electrons as quantum bits (qubits), nevertheless, they are many orders of magnitude shorter than the limit of $T_1$ due to the presence of additional decoherence mechanisms. One mechanism is related to the presence of $^{29}$Si isotopes with non-zero nuclear magnetic moment (natural silicon contains about 4.7% of $^{29}$Si). Dipole-driven flip-flops of $^{29}$Si nuclear spin pairs (termed nuclear spin diffusion) are sensed by the donor-electron spin as random field fluctuations, thus driving decoherence of the electron spin.[17,18] A solution to this problem is to use isotopically-enriched $^{28}$Si with a reduced abundance of $^{29}$Si. Here, we use very high-purity $^{28}$Si crystals with only 50 ppm residual $^{29}$Si, available through the Avogadro project.[19] At 50ppm $^{29}$Si the nuclear spin diffusion processes are largely suppressed (on a time scale shorter than 1 s),[17,18] and therefore other $T_2$ processes can dominate.

Measurements of $T_2$ using 50 ppm $^{29}$Si material with a P donor density of $10^{14}$ /cm$^3$ are shown in Figure 1 (red dots), measured used a standard Hahn echo sequence (90° – τ – 180° – τ – *echo*). For comparison, earlier measurements using 800 ppm $^{29}$Si material with higher donor densities ($10^{15-16}$ /cm$^3$) are also shown (blue and green dots), supporting the observation that donor electron $T_2$ scales inversely with the donor density. Although this data appears to show low-temperature limit of 20 ms for the lowest doped ($10^{14}$ /cm$^3$), this is due to an artifact of the measurement process, known as *instantaneous diffusion*. This effect can be overcome (as described below), leading to longer measured $T_2$ times, indicated by the stars in Figure 1. This extends the temperature range in which $T_2$ is bounded by $T_1$, but it is clear there are other decoherence mechanisms dominant at lower temperatures, which we will demonstrate are related to dipolar interactions between the central donor electron and spins of neighboring donor electrons.

Instantaneous diffusion describes decoherence of observed spins caused by flips of other dipole-coupled electron spins in the bath, as a result of the applied microwave pulses. This is clearly manifested in a dependence of the measured $T_2$ on the rotation angles of the microwave pulses in a Hahn echo experiment, as the effect is suppressed by using small rotation angle pulses in the refocusing pulse.[14,20,21] Figure 2 shows two-pulse echo decays measured at 2.1 K using a $^{28}$Si crystal with a phosphorus density of $1.2 \cdot 10^{14}$ /cm$^3$. In a standard Hahn echo experiment (Figure 2a) with a 180° rotation angle of the second pulse ($\theta_2=180°$), the decay is purely exponential (no spectral diffusion from $^{29}$Si nuclei) and determined by instantaneous diffusion, giving a $T_2$ of 20 ms. On the other hand, when using $\theta_2=14°$ for the refocusing pulse (Figure 2b), the instantaneous diffusion is mostly suppressed and the echo decay is longer, with $T_2 = 0.45$ s.

Completely removing instantaneous diffusion would require using infinitely small rotation angles, $\theta_2$. However, that is not possible because the echo signal intensity also scales to zero as $\theta_2$ decreases. An alternative approach is demonstrated in Figure 3a where we plot the measured $1/T_2$ as a function of $\sin^2(\theta_2/2)$ and then extrapolate the observed linear dependence to $\theta_2 = 0$ (the linear dependence is expected because of the uniform distribution of donors in silicon crystals).[21] The three curves shown in Figure 3a are for three different temperatures in the range 1.8 to 6 K. The slopes of the linear fits are identical at all three temperatures and match the known donor density $1.2 \cdot 10^{14}$ /cm$^3$ in the sample. The vertical intercept (at $\theta_2 = 0$) provides an estimate of the "intrinsic" $T_2$ that would be observed in the absence of instantaneous diffusion effects. It is seen that the intercept decreases as temperature decreases, corresponding to an increase in the intrinsic $T_2$. We note that the linear extrapolation in Figure 3a also removes an additional decoherence mechanism that of dipolar flip-flops between the central spin and a neighbor spin (Figure 3f); the small $\theta_2$ angle makes it unlikely that both spins are flipped, and so the effect of this dipolar interaction is refocused. This has been termed the *direct* flip-flop process,[22] and we will see below how a value for the decoherence rate of this mechanism can be obtained.

The intrinsic $T_2$ measured for three donor densities are plotted as a function of temperature in Figure 3b showing a dependence both on temperature and donor density. Three temperature regions can be identified in Figure 3b. Above 8 K, $T_2$ follows $T_1$ for all three donor densities (dashed line in Figure 3b). The $T_1$ processes (Figure 3c) dominate donor decoherence in the high temperature range. Below 4 K, $T_2$ becomes temperature independent and saturates at a level which is inversely proportional to the donor density. As we show below, $T_2$ in this range is determined by spectral diffusion arising from electron spin flip-flops of nearby donor pairs (Figure 3e), which has been called the *indirect* flip-flop process.[23] At intermediate temperatures (between 4 and 8 K) there appears to be a transitional behavior between the two extremes, however, we find that a simple sum of the two rates from the high and low temperature processes does not provide a good description (dotted lines in Figure 3b). Instead a third decoherence process must be involved, which we identify as spectral diffusion caused by $T_1$-induced spin flips of neighboring donors (Figure 3d).[24,25] A combination of all three processes fully explains the observed temperature dependence of $T_2$ for the donor densities shown in Figure 3b (solid lines).

Both spectral diffusion processes illustrated in Figure 3(d-e) are related to random fluctuations of dipole fields from neighbor donor spins decohering the central spin. However the cause of the fluctuations is different in these two processes. In one case spin-lattice relaxation ($T_1$) leads to random flipping of neighboring donor spins, and in the other case dipole-dipole interactions drive spin flip-flops in neighboring donor pairs. The theory of the first process, termed $T_1$-type spectral diffusion, has been developed previously,[20,24,25] predicting a non-exponential echo decay of the form $\exp[-(2\tau/T_{SD})^2]$, with $T_{SD}^2 \sim T_1/[P]$. In the Supplementary Information we use the measured $T_1$ and the known donor density [P] to demonstrate that the donor two-pulse echo decays measured at 4-8 K are well described by this $T_1$-type spectral diffusion without adjustable parameters. However, below 4 K, where the donor spin $T_1$ becomes extremely long and $T_1$-induced spin flips very rare, this process no longer contributes significantly to donor decoherence.

Two experimental observations suggest that electron spin flip-flopping (Figure 3e) in neighbor pairs is the dominant decoherence process at temperatures below 4 K: (1) $T_2$ shows no temperature dependence, and (2) $T_2$ scales with the donor density. Flip-flopping is driven by dipolar interactions and requires that the interactions be greater than the difference in resonance frequencies ($\Delta\nu = \nu_1 - \nu_2$) of the two spins involved.[3,26] For donor densities $1.2 \cdot 10^{14}$ – $3.3 \cdot 10^{15}$/cm$^3$, as in our samples, the average donor separation is 85-250 nm, and therefore an average spin flip-flopping rate in donor pairs is about 2-40 Hz. The importance of inhomogeneous fields and their role in suppressing donor spin flip-flopping has been recently discussed by Witzel et al.[23] Assuming a donor density of $1.2 \cdot 10^{14}$/cm$^3$, as in one of our $^{28}$Si samples, and taking into account an inhomogeneous broadening from 50 ppm $^{29}$Si, their estimate of the donor $T_2 \sim 1$ second is in agreement with the results in Fig. 3b. Remarkably, below about 8 K, the donor electron is sensitive to interactions between donors which are ~200 nm away. This has important implications for the design of donor qubit architectures and their fabrication using for example, ion-implantation to create arrays of interacting donors. It also suggests that donor electron spins may also be useful as local spin probes.[27]

It is possible to control the effect of these long-range interactions and inhibit spin flip-flops by artificially increasing the offset ($\Delta\nu$) in resonant frequencies between nearby donors by applying an external magnetic field gradient. The effect of a 1 Gauss/cm magnetic field gradient is shown in Figure 4. The magnitude of the gradient was estimated from the increase in the ESR linewidth, from 26 mG to 0.22 G (Figure 4b), together with the dimensions (2x2x8 mm$^3$) of our sample. The intercept after extrapolating to $\theta_2 = 0$ corresponds to an increase of the intrinsic $T_2$ from 1.3 ± 0.1 s in the absence of a gradient to $T_2 \sim 10$ s in the presence of the gradient. A gradient-induced increase in the intrinsic $T_2$ was also observed in other $^{28}$Si crystals with higher donor densities.

As seen in Figure 4a, the slope of $T_2$ vs $\sin^2(\theta_2/2)$ also decreases upon applying the gradient. Both changes in the intercept and the slope can be understood in terms of suppressing spin flip-flops: the intercept changes due to suppression of the indirect flip-flops (Figure 3e) and the slope changes due to suppression of the direct flip-flops (Figure 3f). Individual contributions of direct and indirect flip-flop mechanisms can be extracted from a simultaneous fit of both (gradient and no-gradient) data sets using the expression:[22]

$$\frac{1}{T_2} = \sin^2(\theta_2/2) \cdot \left[ \frac{1}{T_{2(ID)}} + \frac{1}{T_{2(dff)}} S_{ff} \right] + \frac{1}{T_{2(iff)}} S_{ff} .$$

Here we recognize that both the instantaneous diffusion ($T_{2(ID)}$) and direct flip-flop ($T_{2(dff)}$) processes scale as a function of $\sin^2(\theta_2/2)$, and the indirect flip-flop ($T_{2(iff)}$) process is independent of $\theta_2$. We also introduce a flip-flop suppression factor $S_{ff}$ (when the gradient is applied) and we assume this factor to be the same for both direct and indirect flip-flop processes. The fits, shown in Figure 4, give $T_{2(iff)} = 1.3 \pm 0.1$ s and $T_{2(dff)} = 0.8 \pm 0.15$ s, while $S_{ff} = 13 \pm 8$ %.

The 1 Gauss/cm gradient introduces a shift of the resonant frequencies of $\Delta\nu \sim 100$ Hz for donors at a separation of 250 nm (corresponding to $1.2 \cdot 10^{14}$/cm$^3$). Using the flip-flop suppression factor $S_{ff} = 13$ % induced by this gradient, we then estimate the intrinsic distribution of donor resonance frequencies in the crystal to be $\Delta\nu \sim 16$ Hz before applying the gradient (see Supporting Information). This value is a rough estimate and intended only to provide a qualitative explanation of the gradient effect. It is lower than what is expected from the random configurations of 50 ppm $^{29}$Si nuclei using Kittel-Abrahams result,[28] which predicts $\Delta\nu \sim 280$

Hz; the discrepancy may be due to the fact that the 50 ppm abundance is far into a low-concentration limit where the Kittel-Abrahams formula is not accurate.

To conclude, we have demonstrated that the $T_2$ of electrons spins bound to donors in silicon can be as long as about 10 seconds at 1.8 K. This required the use of very pure $^{28}$Si crystals (to reduce spectral diffusion from $^{29}$Si) and the identification (and subsequent suppression) of three decoherence mechanisms arising from dipolar interactions between donor electron spins. It should be noted that the extrapolation procedure necessary to deal with the effects of instantaneous diffusion will tend to mask decoherence with a non-exponential time dependence. In particular, spectral diffusion from residual $^{29}$Si has the form of an exponential of time raised to a power, and this procedure will be sensitive only to the early-time behavior of that function, since that is when the data points are taken.

The $T_2$ of 10 seconds is still two orders of magnitude shorter than $T_1 = 2,000$ seconds at this temperature, and the remaining decoherence might be related to a residual donor flip-flopping that was not fully suppressed by applying the 1D field gradient, to residual $^{29}$Si, or to other yet undetermined mechanisms. Further work will be required to see if $T_2$ can be pushed to even longer times and to identify the remaining decoherence mechanisms. Methods of further reducing the effect of donor electron spin flip-flops include using $^{28}$Si with lower doping densities (but proportionally smaller signals), increasing the ratio of Zeeman energy to temperature by reducing the temperature and/or increasing the magnetic field, and refocusing the dipolar interactions with MREV-type pulse sequences.[29,30]

We thank W. M. Witzel and A. Morello for helpful discussions. Work at Princeton was supported by the NSF through the Princeton MRSEC (DMR-0213706) and the NSA/LPS through LBNL (MOD 713106A), at Keio by the MEXT, FIRST, NanoQuine, Keio G-COE and JST-EPSRC/SIC (EP/H025952/1), at Oxford by the EPSRC through CAESR (EP/D048559/1), at LBNL by the DOE (DE-AC02-05CH11231) and the NSA under 100000080295, at SFU by NSERC. J. J. L. M. is supported by the Royal Society.

## Methods

High-purity $^{28}$Si-enriched single crystals with phosphorus donor densities ranging from $1.2 \cdot 10^{14}$ to $3.3 \cdot 10^{15}$ /cm$^3$ and a $^{29}$Si concentration of 50 ppm were obtained from a dislocation free single crystal doped using PH$_3$ during floating-zone growth from highly enriched polysilicon.[19] Pulsed EPR experiments were performed using an X-band (9.7 GHz) Bruker EPR spectrometer (Elexsys 580) equipped with a low-temperature helium-flow cryostat (Oxford CF935). Typical π/2 and π pulses were 40 and 80 ns, respectively. For temperatures below 5 K (when T$_1$ relaxation was longer than 10 s), a light emitting diode (1050 nm) was pulsed for 50 ms after each pulsed experiment to promote a faster thermalization of donor spins. Particular care was taken to suppress mechanical (microphonic) vibrations in the cryostat setup, and to reduce the magnetic field noise introduced through pickup in the field-controller circuitry.

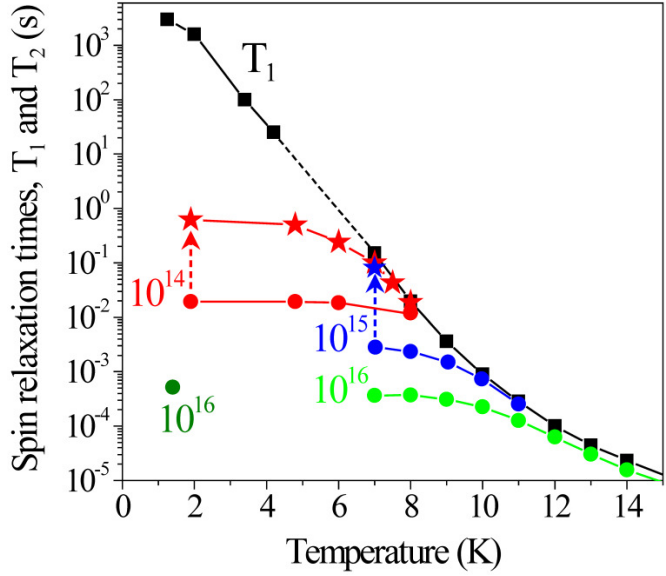

**Figure 1.** Summary of measured spin relaxation times, $T_1$ and $T_2$ for phosphorus donors in silicon at cryogenic temperatures. The longitudinal spin relaxation time $T_1$ (black squares) changes by eight orders of magnitude in the temperature range from 1.2 to 15 K, limited by one-phonon (direct) and two-phonon (Raman) relaxation processes below 4 K[11] and by an Orbach relaxation process above 7 K.[12,14] In contrast to $T_1$, the figure shows that the transverse spin relaxation time $T_2$ demonstrates a substantial dependence on donor density. The $T_2$ data (filled circles marked with respective donor densities) were taken from the current work, as well as Refs. [13,14]. $T_2$ is bounded by $T_1$ at high temperatures but then saturates at low temperatures at a level inversely proportional to donor density due to dipole interaction between donors (instantaneous diffusion). By suppressing instantaneous diffusion the longer (intrinsic) $T_2$ can be revealed (stars), limited by $T_1$ processes at 8 K and above, and by dipole interactions with neighboring donors below 8 K. The longest $T_2 = 0.6$ s measured is still more than 3 order of magnitude shorter than its fundamental limit $T_1 \sim 2000$ s at 1.8 K.

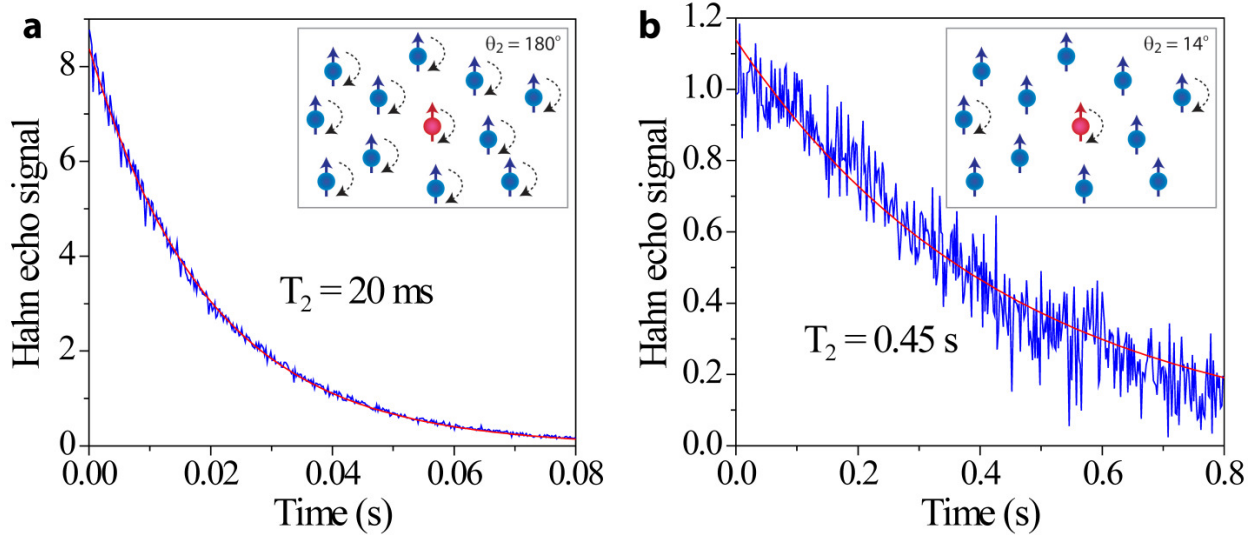

**Figure 2.** Electron spin echo decays of phosphorus donors in $^{28}$Si crystal with 50 ppm $^{29}$Si. The two-pulse echo sequence ($90° - \tau - \theta_2 - \tau - echo$) was used in both cases, with the rotation angle of the second pulse set to (a) $\theta_2 = 180°$ and (b) $\theta_2 = 14°$. The relaxation decay is short in (a), with $T_2 = 20$ ms being totally determined by instantaneous diffusion. When using a small $\theta_2 = 14°$ (b), the instantaneous diffusion is mostly suppressed, revealing a much longer $T_2 = 0.45$ s. Red curves are exponential fits. Donor concentration was $1.2 \cdot 10^{14}$/cm$^3$, and temperature 2.1 K. (Inserts) When using $\theta_2 = 180°$, all neighbor spins (blue) are flipped by the second pulse resulting in a large net change of dipolar interactions as seen by the central spin (red) which leads to a strong instantaneous diffusion. Only few neighbor spins are flipped when $\theta_2 = 14°$, and therefore the instantaneous diffusion is strongly suppressed.

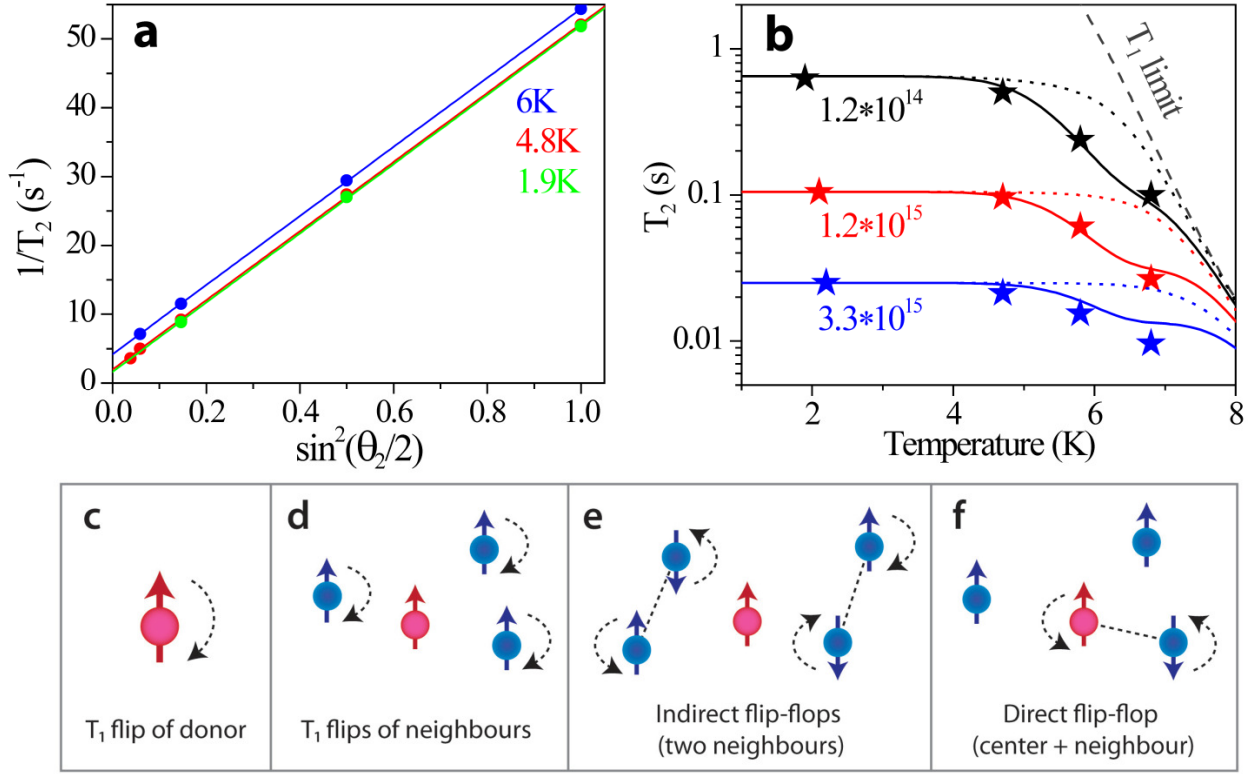

**Figure 3.** "Intrinsic" $T_2$ obtained upon suppressing instantaneous diffusion. (a) The experimental $1/T_2$ rates are plotted as a function of a rotation angle $\theta_2$, measured at three different temperatures. The slopes of the linear fits correlate with the known donor density of $1.2 \cdot 10^{14}/\text{cm}^3$ in the $^{28}$Si crystal. Intercepts obtained by extrapolating to zero $\theta_2$ give an "intrinsic" $T_2$ corresponding to fully suppressing instantaneous diffusion. (b) The intrinsic $T_2$ (stars) are plotted as a function of temperature for three $^{28}$Si crystals with different donor densities as indicated. Solid lines are fits assuming three relaxation mechanisms: (c) $T_1$ relaxation of a central spin, (d) $T_1$-induced flips of neighboring donor spins, and (e) spin flip-flops in neighboring donor pairs. The dotted lines show a much poorer fit assuming only two relaxation processes (c) and (e). The dashed curve indicates the $T_1$ relaxation limit. Cartoons (c-f) illustrate four relaxation mechanisms discussed in the text, including those related to dipolar interactions to neighbor donors. Flips and flip-flops of neighbor donors (blue) produce fluctuating dipolar fields at a central spin (red) and thus dephase it.

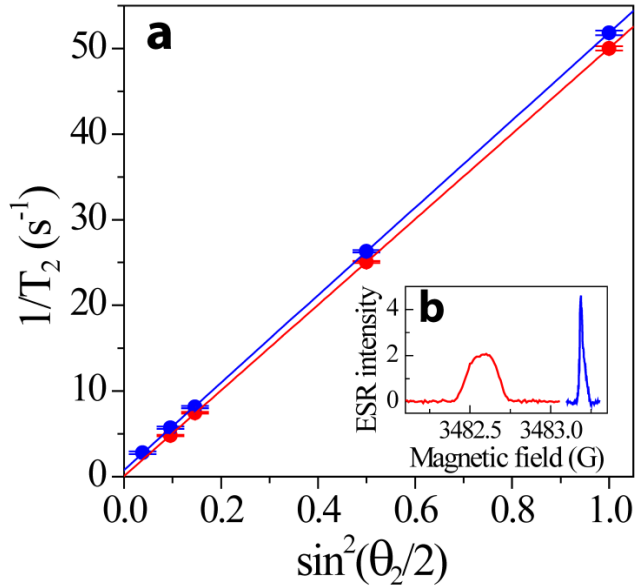

**Figure 4.** Applying an external magnetic field gradient suppresses donor flip-flops and leads to an extended $T_2$. (a) The experimental $1/T_2$ rates are plotted as a function of a rotation angle $\theta_2$, as measured in the absence (blue dots) and presence (red dots) of an external field gradient 1 Gauss/cm. Error bars are shown for each point. The field gradient causes a drop in the $\theta_2 = 0$ intercept of the linear fit, such that the estimated intrinsic $T_2$ increases to ~10 s. The insert (b) illustrates the increase of ESR linewidth from 30 mG (blue trace) to 220 mG (red trace) upon applying the field gradient. Donor concentration was $1.2 \cdot 10^{14}/cm^3$, and temperature 1.8 K.

# Supplementary Information:

# Electron spin coherence exceeding seconds in high purity silicon

Alexei M. Tyryshkin[1], Shinichi Tojo[2], John J. L. Morton[3], Helge Riemann[4], Nikolai V. Abrosimov[4], Peter Becker[5], Hans-Joachim Pohl[6], Thomas Schenkel[7], Mike L. W. Thewalt[8], Kohei M. Itoh[2], and S. A. Lyon[1]

## 1. Spin decoherence by spectral diffusion due to $T_1$-induced spin flips of neighbor donors at intermediate temperatures 4-8 K.

Spin flips of neighbor donors caused by $T_1$ relaxation processes are seen by the central spin as random dipolar field fluctuations (Figure 3d in the main text). These field fluctuations lead to decoherence of the central spin, the process known as $T_1$-type spectral diffusion. In a two-pulse echo experiment, $(90° - \tau - \theta_2 - \tau - \text{echo})$, this spectral diffusion process results in a non-exponential echo signal decay.[1,2] In the limit of $T_1 \gg \tau$ (inter-pulse delay), the decay follows a quadratic exponential dependence:

$$\exp[-(2\tau/T_{SD})^2], \qquad (1)$$

with characteristic spectral diffusion time $T_{SD}^2 = \frac{18\sqrt{3}}{\mu_0} \cdot \frac{\hbar}{(g\beta_e)^2} \cdot \frac{T_1}{[P]}$, where $\mu_0$ vacuum permeability, $g$ electron g-factor, $\beta_e$ Bohr magneton, $\hbar$ Plank constant, and $[P]$ donor density.

Figure A1 shows the two-pulse echo decays measured for $^{28}$Si crystal with donor density $1.2 \cdot 10^{14}$/cm$^3$ at three different temperatures. The small rotation angle $\theta_2 = 22$ deg was used to suppress otherwise dominating instantaneous diffusion effects. The solid lines are theoretically predicted decays according Eq. (1). We note that there are no fit parameters and the decay were calculated using the known donor density in the sample and the measured $T_1$ times at each temperature. The simulated decays correlate closely with the experimental curves at 4.7 and 5.8 K, revealing that $T_1$-type spectral diffusion is the dominant decoherence process at these two temperatures. On the other hand, the predicted decay is much slower than seen in the experiment at 2.1 K. The $T_1$-type spectral diffusion makes a negligible contribution at this low temperature, and some other processes dominate the relaxation. It is shown in the main text that this other process is $T_2$-type spectral diffusion due to spin flip-flops in neighbor donor pairs.

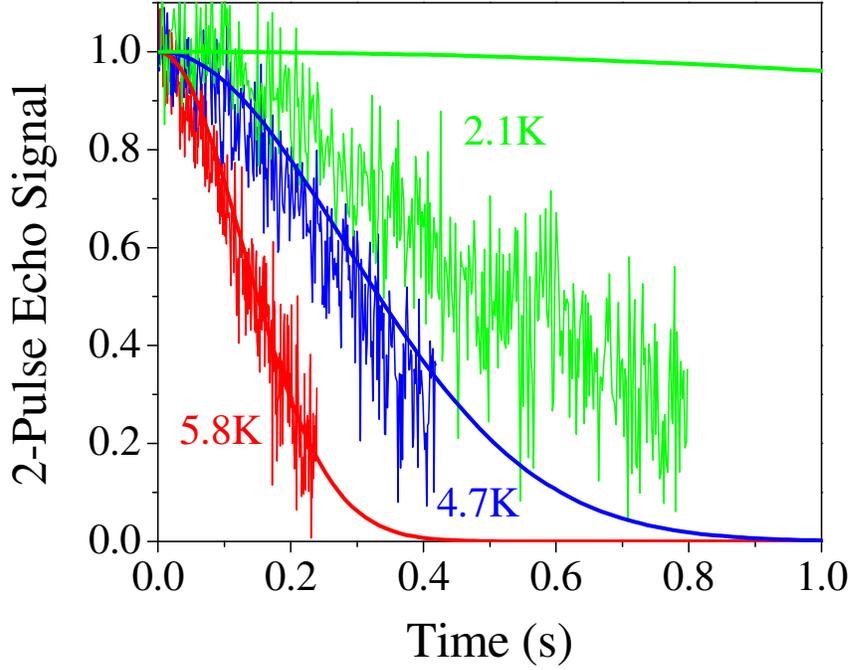

Figure A1. Experimental and simulated two-pulse echo decays of phosphorus donors in $^{28}$Si doped with $1.2 \cdot 10^{14}$ donors/cm$^3$ at three temperatures as indicated. The experimental decays were measured with a pulse sequence ($90° - \tau - \theta_2 - \tau -$ echo) using a small rotation angle $\theta_2$ = 22 deg to suppress instantaneous diffusion. The simulated decays were calculated using Eq.(1) and assuming $T_1$ = 1.6 s (5.8 K), 8 s (4.7 K), and 1300 s (2.1 K) as measured in the separate experiments.

**2. Intrinsic distribution of resonance offsets ($\Delta \nu = \nu_1 - \nu_2$) between donors in pairs in a $^{28}$Si crystal without an external field gradient.**

In the main text (Figure 4a) we showed that application of an external field gradient suppresses flip-flops in donor pairs leading to longer donor $T_2$. For the gradient of 1 Gauss/cm applied to a $^{28}$Si crystal with $1.2 \cdot 10^{14}$ donors/cm$^3$ we estimated the flip-flop suppression factor to be $S_{ff}$ = 13 %.

Flip-flops in donor pairs are driven by dipolar interactions and also controlled by local inhomogeneous fields. For a flip-flop to occur requires that the dipolar interaction ($a_{dd}$) within a pair be greater than the inhomogeneous offset of resonant frequencies ($\Delta \nu = \nu_1 - \nu_2$) of two spins involved. Thus, a pair can flip-flop only if $a_{dd} > \Delta \nu$, and the pair cannot flip-flop if $a_{dd} < \Delta \nu$. By applying a known external field gradient one can controllably increase $\Delta\nu$, to make it greater than $a_{dd}$, and thus to switch off flip-flops in those pairs that were flip-flopping before applying the gradient.

For donor density of $1.2 \cdot 10^{14}$ donors/cm$^3$ the average distance between donors is 255 nm and the average dipolar interaction is $a_{dd} \sim 2$ Hz. Donor resonance offsets ($\Delta\nu$) in a $^{28}$Si crystal are described by an unknown distribution that arises from various inhomogeneous fields, including $^{29}$Si hyperfine fields from random configurations of $^{29}$Si nuclei, local crystal strains from various impurities and defects, and etc. The shape of this distribution is predicted to be Lorentzian[3] and the width of the distribution is unknown. The question then arises whether we can estimate the width by using the known flip-flop suppression factor $S_{ff}$ = 13 % as determined for the gradient 1Gauss/cm.

Figure A2 shows the simulated $\Delta\nu$ distributions. We assumed a Lorentzian inhomogeneous linewidth of $\Delta\nu \sim 15$ Hz in the absence of a gradient. The 1 G/cm gradient was assumed orientated at 45 degrees with respect to the external magnetic field, $B_0$, as it was in the experiment, and then the offset distribution was calculated for the average donor-donor distance of 255 nm (corresponding to $1.2 \cdot 10^{14}$ donors/cm$^3$). It is seen that the offset distribution transforms significantly upon applying the gradient. In both cases the average $a_{dd} \sim 2$ Hz (at $1.2 \cdot 10^{14}$ donors/cm$^3$) is smaller than the width of the distribution. Only pairs with resonant offsets less than 2 Hz can flip-flop, and the integrated number of flip-flopping pairs changes from 5 % in the absence of a gradient to 0.7 % in the presence of a 1G/cm gradient. Their ratio corresponds to a flip-flop suppression factor of 14 % which is close to that found in the experiment.

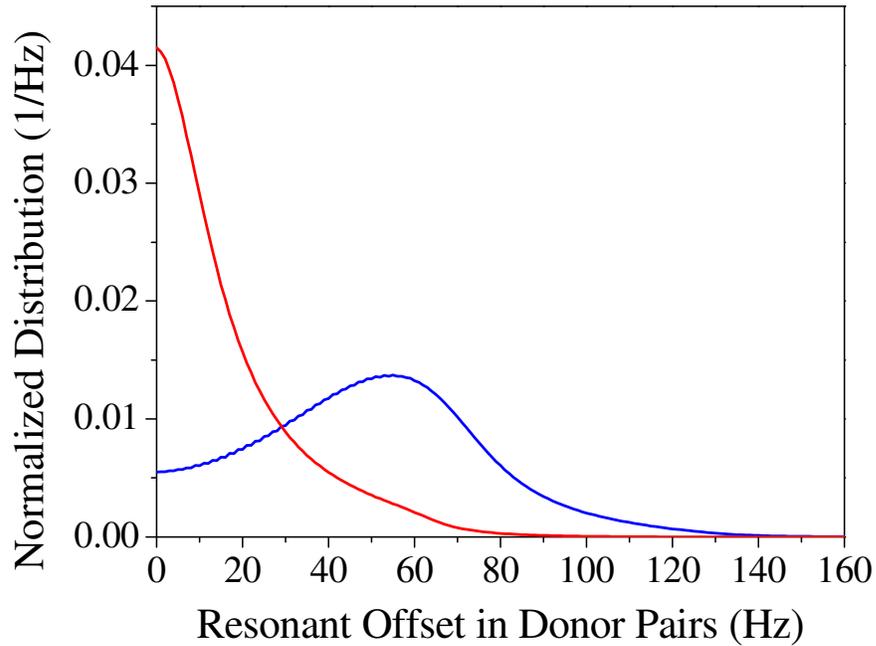

Figure A2. Distribution of resonance offsets ($\Delta \nu = \nu_1 - \nu_2$) between donors in pairs at concentration $1.2 \cdot 10^{14}$ donors/cm$^3$, in the absence (red) and presence (blue) of the externally applied field gradient of 1 Gauss/cm. The gradient is assumed to be oriented at 45 degree with respect to the external magnetic field, $B_0$.